\documentclass[preprint,showpacs,preprintnumbers,amsmath,amssymb,prd]{revtex4}
\usepackage{graphicx}
\newcommand{\del}[1]{ \partial_{#1} }

\def\al{{\alpha}}
\def\lam{{\lambda}}
\def\gam{{\gamma}}

\def\tilU{{\tilde{U}}}
\def\tU{{\tilde{U}}}
\def\mE{{{\mathcal E}}}
\begin{document}
\title{Black di-ring and infinite nonuniqueness}

\author{Hideo Iguchi and Takashi Mishima} 
\affiliation{
Laboratory of Physics,~College of Science and Technology,~
Nihon University,\\ Narashinodai,~Funabashi,~Chiba 274-8501,~Japan
}
\date{\today}

\begin{abstract}
We show that the $S^1$-rotating black rings can be superposed by the
solution generating technique.
 We analyze the black di-ring solution for the simplest case of multiple
rings. 
There exists an equilibrium black di-ring where
the conical singularities are cured by the suitable 
choice of physical parameters.
Also there are infinite numbers of black di-rings with the same mass and 
angular momentum. These di-rings can have two different continuous 
limits of single black rings. Therefore we can transform the fat black
ring to the thin ring 
with the same mass and angular momentum by way
of the di-ring solutions.
\end{abstract}
\maketitle

\section{Introduction}

One of the most important recent findings of the higher-dimensional 
general relativity is a single-rotational black ring solution by Emparan and Reall
\cite{ref1}. (See also \cite{Emparan:2006mm}.)
This solution is a vacuum, axially symmetric and asymptotically flat solution
of the five-dimensional general relativity.
The topology of the event horizon is $S^1 \times S^2$. The black ring rotates
along the direction of $S^1$. 
The balanced black ring which has no conical singularity
has a minimum of angular momentum for a fixed mass
parameter. When the angular momentum is near this minimum,
there are two different black rings with the
same angular momentum. They are called fat and thin black rings according to 
their shapes. In addition we have a single-rotational spherical
black hole \cite{Myers:1986un} with the same asymptotic parameters. 
This finding entails the discrete
nonuniqueness of the five dimensional vacuum solutions.
It has been shown that the black rings can have dipole charges
which are independent of all conserved charges
\cite{Emparan:2004wy}. Therefore the dipole rings imply the infinite
violation of uniqueness by a continuous parameter.


Recently the present authors found a black ring solution with $S^2$ rotation
by using a solitonic solution-generating technique
 \cite{{Mishima:2005id},{Iguchi:2006tu}}.
The seed solution of this ring is a simple Minkowski spacetime.
Because the effect of rotation cannot compensate for 
the gravitational attractive force, the ring has a kind of strut structure.
We have also generated the black ring with $S^1$ rotation by 
the same solitonic solution-generating technique \cite{Iguchi:2006rd}.
The seed solution is not a Minkowski spacetime but an Euclidean C-metric solution.
It has been shown that these two solutions can be obtained by the inverse 
scattering method
 \cite{{Azuma:2005az},{Tomizawa:2005wv},{Tomizawa:2006vp},{Tomizawa:2006jz}}.
Rotating dipole black ring solutions have been systematically generated
in five-dimensional Einstein-Maxwell-dilaton gravity
\cite{{Yazadjiev:2006hw},{Yazadjiev:2006ew}}.
The relations between the seed and the solitonic solutions
can be easily understood through the analysis of their rod structures
\cite{{ref8},{refHAR}}.
The seed of $S^1$-rotating black ring has been constructed by the help of 
 the rod structure analysis.
Thus the rod structure analysis is expected to be a useful guide
to construct seed solutions for new solutions.

In this paper, we consider the multiplexed $S^1$-rotating black rings
arranged in a concentric pattern.
The seed solution can be constructed by the help of rod structure analysis
as in the case of the $S^1$-rotating black ring \cite{Iguchi:2006rd}.
The exact expressions of metric functions can be written down
by the solitonic transformation, but in rather complicated forms.
In this paper we analyze di-ring solutions as the first step in the black ring
multiplication.
In the supersymmetric system the solution of multiple black rings exist indeed
\cite{{Gauntlett:2004wh},{Gauntlett:2004qy}}.
Also the solution of concentric static extremal black rings has been
considered \cite{Teo:2005wf}.

The simplest multiple black rings solution is a black di-ring. 
This solution has two ring-like 
event horizons of different radii with the same topology 
of $S^1 \times S^2$. Both horizons can rotate along the direction of $S^1$.
As similar as the single black ring solutions, 
this solutions has conical singularities
for general values of parameters. However these conical singularities 
can be cured by an appropriate choice of the parameters as in the case of the
$S^1$-rotating black ring. The black di-rings can have the same mass and angular 
momentum for infinite numbers of sets of parameters.
Therefore the black di-rings realize an infinite nonuniqueness 
without dipole charges.
There can exist one Myers-Perry black hole,
two $S^1$-rotating black rings and infinite numbers of black di-rings
for the same mass and angular momentum. 
Also these single black rings are continuous limits in the black di-rings.
Therefore these two different black rings are connected by 
the black di-rings with the same mass and angular momentum.
When we shrinks the inner ring down to zero radius, we obtain a solution 
describing a black hole sitting at the common center of the outer ring.

The plan of the paper is as follows.  In Sec. \ref{sec:technique}
we briefly review the solution-generating technique used in the analysis.
The rod structure analysis is explained in Sec. \ref{sec:rods}.
We give the seed solutions of black multi- and di-ring solutions and analyze the some features of di-ring solution in Sec. \ref{sec:multi}. 
In Sec. \ref{sec:summary} we give a summary of this article.

\section{Brief review of solution generating technique}
\label{sec:technique}

At first we briefly explain the procedure to generate axisymmetric solutions
in the five-dimensional general relativity. 
The spacetimes which we considered 
satisfy the following conditions:
(c1) five dimensions, (c2) asymptotically flat spacetimes, 
(c3) the solutions of 
vacuum Einstein equations, (c4) having three commuting Killing vectors 
including time translational invariance and 
(c5) having a single non-zero angular momentum component. 
Under the conditions (c1) -- (c5), 
we can employ the following Weyl-Papapetrou metric form 
\begin{eqnarray}
ds^2 &=&-e^{2U_0}(dx^0-\omega d\phi)^2+e^{2U_1}\rho^2(d\phi)^2
       +e^{2U_2}(d\psi)^2 
       +e^{2(\gamma+U_1)}\left(d\rho^2+dz^2\right) ,
       \label{WPmetric}
\end{eqnarray}
where $U_0$, $U_1$, $U_2$, $\omega$ and $\gamma$ are functions of 
$\rho$ and $z$. 
Then we introduce new functions 
$S:=2U_0+U_2$ and $T:=U_2$ so that 
the metric form (1) is rewritten into 
\begin{eqnarray}
ds^2 &=&e^{-T}\left[
       -e^{S}(dx^0-\omega d\phi)^2
       +e^{T+2U_1}\rho^2(d\phi)^2  
+e^{2(\gamma+U_1)+T}\left(d\rho^2+dz^2\right) \right]
  +e^{2T}(d\psi)^2.
  \label{MBmetric}
\end{eqnarray}
Using this metric form
the Einstein equations are reduced to the following set of equations, 
\begin{eqnarray*}
&&{\bf\rm (i)}\quad
\nabla^2T\, =\, 0,   \\
&&{\bf\rm (ii)}
\left\{\begin{array}{ll}
& \del{\rho}\gamma_T={\displaystyle
  \frac{3}{4}\,\rho\,
  \left[\,(\del{\rho}T)^2-(\del{z}T)^2\,\right]}\,\ \   \\[3mm]
& \del{z}\gamma_T={\displaystyle 
\frac{3}{2}\,\rho\,
  \left[\,\del{\rho}T\,\del{z}T\,\right],  }
 \end{array}\right.  \\
&&{\bf\rm (iii)}\quad
\nabla^2\mE_S=\frac{2}{\mE_S+{\bar\mE}_S}\,
                    \nabla\mE_S\cdot\nabla\mE_S , \\  
&&{\bf\rm (iv)}
\left\{\begin{array}{ll}
& \del{\rho}\gamma_S={\displaystyle
\frac{\rho}{2(\mE_S+{\bar\mE}_S)}\,
  \left(\,\del{\rho}\mE_S\del{\rho}{\bar\mE}_S
  -\del{z}\mE_S\del{z}{\bar\mE}_S\,
\right)}     \\
& \del{z}\gamma_S={\displaystyle
\frac{\rho}{2(\mE_S+{\bar\mE}_S)}\,
  \left(\,\del{\rho}\mE_S\del{z}{\bar\mE}_S
  +\del{\rho}\mE_S\del{z}{\bar\mE}_S\,
  \right)},  
\end{array}\right.  \\
&&{\bf\rm (v)}\quad
\left( \del{\rho}\Phi,\,\del{z}\Phi \right)
=\rho^{-1}e^{2S}\left( -\del{z}\omega,\,\del{\rho}\omega \right),  \\
&&{\bf\rm (vi)}\quad 
\gamma=\gamma_S+\gamma_T,   \\
&&{\bf\rm (vii)}\quad 
U_1=-\frac{S+T}{2},
\end{eqnarray*}
where $\Phi$ is defined through the equation (v) and the function 
$\mathcal{E_S}$ is defined by 
$
\,\mE_S:=e^{S}+i\,\Phi\,.
$
The most non-trivial task to obtain new metrics is to solve 
the equation (iii) because of its non-linearity. 
To overcome this difficulty 
here we use the method similar to the Neugebauer's 
B\"{a}cklund transformation \cite{{Neugebauer:1980}}
or the Hoenselaers-Kinnersley-Xanthopoulos transformation \cite{Hoenselaers:1979mk}.

To write down the exact form of the metric functions,
we follow the procedure
given by Castejon-Amenedo and Manko \cite{ref7}.
In the five dimensional spacetime we start from the following form of a seed static metric
\begin{eqnarray}
ds^2 &=& e^{-T^{(0)}}\left[
       -e^{S^{(0)}}(dx^0)^2
       +e^{-S^{(0)}}\rho^2(d\phi)^2 
   +e^{2\gamma^{(0)}-S^{(0)}}\left(d\rho^2+dz^2\right) \right]
  +e^{2T^{(0)}}(d\psi)^2.
\nonumber
\end{eqnarray}
For this static seed solution, $e^{S^{(0)}}$, of the Ernst equation (iii), 
a new Ernst potential can be written in the form
\begin{equation}
{\cal E}_S = e^{S^{(0)}}\frac{x(1+ab)+iy(b-a)-(1-ia)(1-ib)}
                         {x(1+ab)+iy(b-a)+(1-ia)(1-ib)},
\nonumber
\end{equation}
where $x$ and $y$ are the prolate spheroidal coordinates:
$
\,\rho=\sigma\sqrt{x^2-1}\sqrt{1-y^2},\ z=\sigma xy\,,
$
with $\sigma>0$. The ranges of these coordinates are 
$1 \le x$ and $-1 \le y \le 1$.
The functions $a$ and $b$ satisfy the following 
simple first-order differential equations 
\begin{eqnarray}
(x-y)\del{x}a&=&
a\left[(xy-1)\del{x}S^{(0)}+(1-y^2)\del{y}S^{(0)}\right], \nonumber \\
(x-y)\del{y}a&=&
a\left[-(x^2-1)\del{x}S^{(0)}+(xy-1)\del{y}S^{(0)}\right], \nonumber \\
(x+y)\del{x}b&=&
-b\left[(xy+1)\del{x}S^{(0)}+(1-y^2)\del{y}S^{(0)}\right] , \nonumber\\
(x+y)\del{y}b&=&
-b\left[-(x^2-1)\del{x}S^{(0)}+(xy+1)\del{y}S^{(0)}\right]. \nonumber \\
\label{eq:ab}
\end{eqnarray}
For the typical seed
\begin{equation}
 S^{(0)}= \frac{1}{2} \ln [R_d+(z-d)],
\end{equation}
the following $a$ and $b$ satisfy the 
differential equations (\ref{eq:ab}),
\begin{equation}
a=l_{\sigma}^{-1}e^{2\phi_{d,\sigma}}\ ,\ \ \ 
b=-l_{-\sigma}e^{-2\phi_{d,-\sigma}}\ ,\label{ab_phi}
\end{equation}
where
\begin{equation}
\phi_{d,c}=\frac{1}{2}
 \ln\left[\,e^{-\tilU_{d}}\left(e^{2U_c}+e^{2\tilU_{d}}\right)\,\right].
 \label{phi_i}
\end{equation}
Here the functions $\tilU_{d}$ and $U_{c}$ are defined as $\tilU_{d}:=\frac{1}{2}\ln\left[\,R_{d}+(z-d)\,\right]$ and 
$U_{c}:=\frac{1}{2}\ln\left[\,R_{c}-(z-c)\,\right]$.
Because of the linearity of the differential equations (\ref{eq:ab})
for $S^{(0)}$,
we can easily obtain $a$ and $b$ which correspond to a general seed function.

The metric functions for the five-dimensional metric 
 (\ref{MBmetric}) are obtained
by using the formulas shown by \cite{ref7}, 
\begin{eqnarray}
e^{S}&=&e^{S^{(0)}}\frac{A}{B}   \label{e^S} \\
\omega&=&2\sigma e^{-S^{(0)}}\frac{C}{A}+C_1 \label{omega}     \\
e^{2\gamma}&=&C_2(x^2-1)^{-1}A
                e^{2\gamma'}, \label{e_gamma}
\end{eqnarray}
where $C_1$ and $C_2$ are constants and
$A$, $B$ and $C$ are given by
\begin{eqnarray*}
A&:=&(x^2-1)(1+ab)^2-(1-y^2)(b-a)^2, \\
B&:=&[(x+1)+(x-1)ab]^2+[(1+y)a+(1-y)b]^2, \\
C&:=&(x^2-1)(1+ab)[(1-y)b-(1+y)a]  
\nonumber\\ &&\hskip 0.9cm
+(1-y^2)(b-a)[x+1-(x-1)ab].
\end{eqnarray*}
In addition 
the $\gamma'$ in Eq. (\ref{e_gamma}) is a $\gamma$ function corresponding to the static metric,
\begin{eqnarray}
ds^2 &=& e^{-T^{(0)}}\left[
       -e^{2U^{\mbox{\tiny(BH)}}_0+S^{(0)}}(dx^0)^2
       +e^{-2U^{\mbox{\tiny(BH)}}_0-S^{(0)}}\rho^2(d\phi)^2 \right. 
    \nonumber \\ &&\hskip -0.cm \left.
   +e^{2(\gamma'-U^{\mbox{\tiny(BH)}}_0)-S^{(0)}}\left(d\rho^2+dz^2\right) \right]
  +e^{2T^{(0)}}(d\psi)^2 \label{static_5}
\end{eqnarray}
where ${\displaystyle U_{0}^{\mbox{\tiny(BH)}}=\frac{1}{2}\ln\left( \frac{x-1}{x+1} \right)}$. 
Therefore the function $\gam'$ obeys the following equations,
\begin{equation}
\del{\rho}\gam'=
\frac{1}{4}\rho\left[(\del{\rho}S')^2-(\del{z}S')^2\right]
 +\frac{3}{4}\rho\left[(\del{\rho}T')^2-(\del{z}T')^2\right],
 \label{eq:drho_gamma'}
\end{equation}
\begin{equation}
\del{z}\gam'=
\frac{1}{2}\rho\left[\del{\rho}S'\del{z}S'\right]
 +\frac{3}{2}\rho\left[\del{\rho}T'\del{z}T'\right],
 \label{eq:dz_gamma'}
\end{equation}
where the first terms are contributons from Eq. (iv) and the 
second terms come from Eq. (ii).
Here the functions $S'$ and $T'$ can be read out from Eq. (\ref{static_5}) as
\begin{eqnarray}
S'&=&2\,U^{(BH)}_0+S^{(0)},  \label{eq:S'} \\
T'&=&T^{(0)}. \label{eq:T'}
\end{eqnarray}
To integrate these equations we can use the following fact 
that, the partial differential equations
\begin{eqnarray}
\del{\rho}\gam'_{cd}
    &=&\rho\left[\del{\rho}\tU_{c}\del{\rho}\tU_{d}
            -\del{z}\tU_{c}\del{z}\tU_{d}\right],  \label{drho_gm}\\
\del{z}\gam'_{cd}
    &=&\rho\left[\del{\rho}\tU_{c}\del{z}\tU_{d}
           +\del{\rho}\tU_{d}\del{z}\tU_{c}\right], \label{dz_gm}
\end{eqnarray}
have the following solution, 
\begin{equation}
\gam'_{cd}=\frac{1}{2}\tU_{c}+\frac{1}{2}\tU_{d}-\frac{1}{4}\ln Y_{cd}, \label{gam'}
\end{equation}
where $Y_{cd}:=R_cR_d+(z-c)(z-d)+\rho^2$. The general solution of $\gamma'$
is given by the linear combination of the functions $\gamma'_{cd}$.
And then the function $T$ is equals to $T^{(0)}$ and $U_1$ is given by 
the Einstein equation (vii).

\section{Rod structure analysis}
\label{sec:rods}

We give a brief explanation of the
rod structure analysis erabolated by Harmark \cite{refHAR}.
See \cite{refHAR} for complete explanations.

Here we denote the D-dimensional axially symmetric 
stationary metric as
\begin{equation}
 ds^2 = G_{ij}dx^i dy^j + e^{\nu}(d\rho^2 + dz^2)
\end{equation}
where $G_{ij}$ and $\nu$ are functions only of $\rho$ and $z$
and $i,j=0,1,\dots,D-3$. 
The $D-2$ by $D-2$ matrix field $G$ satisfies the following
constraint
\begin{equation}
 \rho = \sqrt{|\det G|}.
\end{equation}
The equations for the matrix field $G$ 
can be derived from the Einstein equation $R_{ij}=0$ as
\begin{equation}
 G^{-1}\nabla G = (G^{-1} \nabla G)^2,
 \label{eq:har}
\end{equation}
where the differential operator $\nabla$ is 
the gradient in three-dimensional unphysical flat space with metric
\begin{equation}
 d\rho^2 + \rho^2 d\omega^2 + dz^2.
\end{equation}

Because of the constraint $\rho = \sqrt{|\det G|}$,
at least one eigenvalue of $G(\rho,z)$ goes to zero for 
$\rho \rightarrow 0$. However it was shown that
if more than one eigenvalue goes to zero as $\rho \rightarrow 0$,
we have a curvature singularity there. Therefore  we consider
solutions which have only one eigenvalue goes to zero for 
$\rho \rightarrow 0$, except at isolated values of $z$.
Denoting these isolated values of $z$ as $a_1,a_2,\dots,a_N$,
we can divide the $z$-axis into the $N+1$ intervals
$[-\infty,a_1]$,$[a_1,a_2]$,$\dots$,$[a_N,\infty]$,
which is called as rods. These rods correspond to the source
added to the equation (\ref{eq:har}) at $\rho=0$ to
prevent the break down of the equation there.

The eigenvector for the zero eigenvalue of $G(0,z)$
\begin{equation}
 {\bf v}=v^i \frac{\partial}{\partial x^i},
\end{equation}
which satisfies
\begin{equation}
 G_{ij}(0,z) v^i = 0,
\end{equation}
determines the direction of the rod.
If the value of $\frac{G_{ij}v^i v^j}{\rho^2}$
is negative (positive) for $\rho \rightarrow 0$ the rod is called
timelike (spacelike).
Each rod corresponds to the region of the translational or
rotational invariance
of its direction.
The timelike rod corresponds to a horizon.
The spacelike rod corresponds to a compact direction.

\section{$S^1$-rotating black multi-ring}
\label{sec:multi}

\begin{figure}
  \includegraphics[scale=0.55,angle=0]{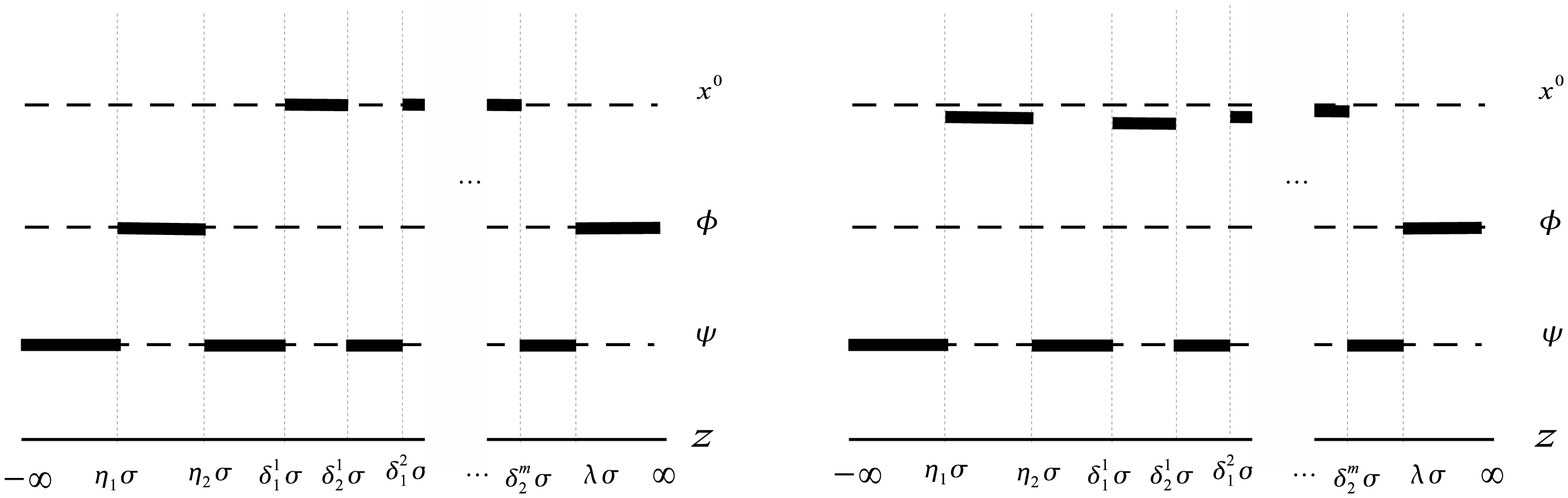}
  \caption{Schematic pictures of rod structures of multi-ring and its seed. 
The left panel
shows the rod structure of seed metric of 
$S^1$-rotating black multi-ring. The right panel shows the rod structure
of $S^1$-rotating black multi-ring. The finite spacelike rod
 $[\eta_1\sigma,\eta_2\sigma]$
in the left panel is altered to the finite timelike rod by the solution-generating transformation.
All static timelike rods may be transformmed to stationary ones by the 
solitonic transformation.
To denote the rotation of the event horizons, we put the finite timelike rods
between the lines of $x^0$ and $\phi$.}
 \label{fig:rods_multi}
 \end{figure}

The $n$-multiplexed $S^1$-rotating black ring can be
obtained in the following manner.
At first we prepare the seed solution of multi-ring as in 
Fig. \ref{fig:rods_multi}.
To assure the asymptotical flatness, we need two semi-infinite spacelike
rod in the different directions.
There is a finite spacelike rod with the direction vector $\partial/\partial \phi$ 
around the $z=0$. Between this finite rod and the semi-infinite spacelike
rod of $\phi$-direction, we alternately arrange $n$ spacelike rods in $\psi$-direction and $(n-1)$ static finite timelike rods.
The finite spacelike rod of $\phi$-direction
is changed to a finite timelike rod with
$\phi$ rotation by the solitonic transformation
 \cite{{Iguchi:2006tu},{Iguchi:2006rd}}.
In addition the finite timelike rods of seed solution
can get the $\phi$ components in their direction vectors through the transformation.

\begin{figure}
  \includegraphics[scale=0.6,angle=0]{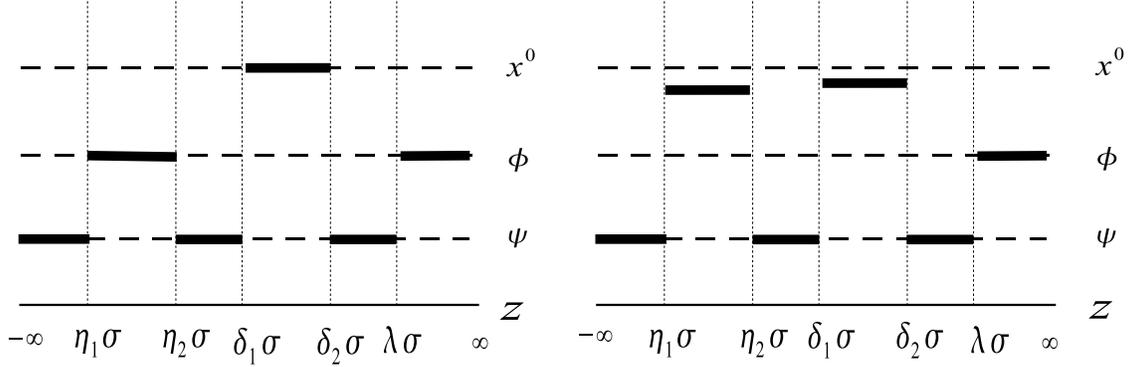}
  \caption{Schematic pictures of rod structures of black di-ring and its seed. 
The left panel
shows the rod structure of seed metric of 
$S^1$-rotating black di-ring. The right panel shows the rod structure
of $S^1$-rotating black di-ring. The finite spacelike rod
 $[\eta_1\sigma,\eta_2\sigma]$
in the left panel is altered to the finite timelike rod by the solution-generating transformation.}
 \label{fig:rods_diring}
 \end{figure}

In the following we investigate the simplest multiple black rings, i.e.,
the black di-ring solution. The rod structure of the seed and the di-ring solution
are given in Fig. \ref{fig:rods_diring}. The rod structure of di-ring
is determined by  
4 length of finite rods and 2 angular velocities of timelike rods.
We have five physical parameters
$\eta_1,\eta_2,\delta_1,\delta_2$ and $\lambda$.
except for the freedom of scaling.
These parameters should satisfy the condition
$-1<\eta_1<\eta_2<1<\delta_1<\delta_2<\lambda$ 
for the di-ring solution.
Note that when we set $\delta_2=\lambda$, the inner ring shrinks to $S^3$ sphere.
When $\delta_1=\delta_2$, these structures are exactly the same as the case of single
$S^1$-rotating black ring.

The seed functions of black di-ring are given by the following functions,
\begin{equation}
T^{(0)}=\tilde{U}_{\lambda \sigma}+\tilde{U}_{\delta_1 \sigma}-\tilde{U}_{\delta_2 \sigma}+\tilde{U}_{\eta_1 \sigma}-\tilde{U}_{\eta_2 \sigma},
\end{equation}
\begin{equation}
S^{(0)}=\tilde{U}_{\lambda \sigma}-(\tilde{U}_{\delta_1 \sigma}-\tilde{U}_{\delta_2 \sigma})+\tilde{U}_{\eta_1 \sigma}-\tilde{U}_{\eta_2 \sigma}.
\end{equation}
The corresponding auxiliary  
potentials of solitonic solutions are obtained as
\begin{equation}
a=\frac{\alpha}{2\sigma^{1/2}} 
 \frac{e^{2U_\sigma}+e^{2\tilU_{\lam\sigma}}}{e^{\tilU_{\lam\sigma}}}
 \frac{e^{\tilU_{\delta_1\sigma}}}{e^{2U_\sigma}+e^{2\tilU_{\delta_1\sigma}}}
 \frac{e^{2U_\sigma}+e^{2\tilU_{\delta_2\sigma}}}{e^{\tilU_{\delta_2\sigma}}}
 \frac{e^{2U_\sigma}+e^{2\tilU_{\eta_1\sigma}}}{e^{\tilU_{\eta_1\sigma}}}
 \frac{e^{\tilU_{\eta_2\sigma}}}{e^{2U_\sigma}+e^{2\tilU_{\eta_2\sigma}}},
\end{equation}
\begin{equation}
b={2\sigma^{1/2}}{\beta}
 \frac{e^{\tilU_{\lam\sigma}}}{e^{2U_{-\sigma}}+e^{2\tilU_{\lam\sigma}}}
 \frac{e^{2U_{-\sigma}}+e^{2\tilU_{\delta_1\sigma}}}{e^{\tilU_{\delta_1\sigma}}}
 \frac{e^{\tilU_{\delta_2\sigma}}}{e^{2U_{-\sigma}}+e^{2\tilU_{\delta_2\sigma}}}
 \frac{e^{\tilU_{\eta_1\sigma}}}{e^{2U_{-\sigma}}+e^{2\tilU_{\eta_1\sigma}}}
 \frac{e^{2U_{-\sigma}}+e^{2\tilU_{\eta_2\sigma}}}{e^{\tilU_{\eta_2\sigma}}},
\end{equation}
where $\alpha$ and $\beta$ are integration constants.
The functions $S'$ and $T'$ in Eqs. (\ref{eq:drho_gamma'}) and
 (\ref{eq:dz_gamma'}) are obtained as
\begin{eqnarray}
S'&=&2\,U^{(BH)}_0+S^{(0)} \nonumber \\
  &=& 2(\tilU_\sigma-\tilU_{-\sigma})
  +\tilde{U}_{\lambda \sigma}
  -(\tilde{U}_{\delta_1 \sigma}-\tilde{U}_{\delta_2 \sigma})
  +\tilde{U}_{\eta_1 \sigma}-\tilde{U}_{\eta_2 \sigma}\\
T'&=&T^{(0)}=\tilde{U}_{\lambda \sigma}+\tilde{U}_{\delta_1 \sigma}-\tilde{U}_{\delta_2 \sigma}+\tilde{U}_{\eta_1 \sigma}-\tilde{U}_{\eta_2 \sigma},
\end{eqnarray}
therfore 
the function $\gamma'$ becomes the following sum of the functions $\gamma'_{cd}$,
\begin{eqnarray}
\gam' &=& \gam'_{\sigma,\sigma}+\gam'_{-\sigma,-\sigma}
+\gam'_{\lambda\sigma,\lambda\sigma}+\gam'_{\delta_1\sigma,\delta_1\sigma}
+\gam'_{\delta_2\sigma,\delta_2\sigma}+\gam'_{\eta_1\sigma,\eta_1\sigma}
+\gam'_{\eta_2\sigma,\eta_2\sigma} \nonumber \\
&& 
 -2\gam'_{\sigma,-\sigma}
+\gam'_{\sigma,\lambda\sigma}-\gam'_{\sigma,\delta_1\sigma}
+\gam'_{\sigma,\delta_2\sigma}+\gam'_{\sigma,\eta_1\sigma}
-\gam'_{\sigma,\eta_2\sigma}
\nonumber \\ &&
-\gam'_{-\sigma,\lambda\sigma}
+\gam'_{-\sigma,\delta_1\sigma}
-\gam'_{-\sigma,\delta_2\sigma}
-\gam'_{-\sigma,\eta_1\sigma}
+\gam'_{-\sigma,\eta_2\sigma}
\nonumber \\ &&
+\gam'_{\lambda\sigma,\delta_1\sigma}
-\gam'_{\lambda\sigma,\delta_2\sigma}
+2\gam'_{\lambda\sigma,\eta_1\sigma}
-2\gam'_{\lambda\sigma,\eta_2\sigma}
-2\gam'_{\delta_1\sigma,\delta_2\sigma}
\nonumber \\ &&
+\gam'_{\delta_1\sigma,\eta_1\sigma}
-\gam'_{\delta_1\sigma,\eta_2\sigma}
-\gam'_{\delta_2\sigma,-\eta_1\sigma}
+\gam'_{\delta_2\sigma,\eta_2\sigma}
-2\gam'_{\eta_1\sigma,\eta_2\sigma} .
\nonumber
\end{eqnarray}
Using these functions we can write down the metric functions of black di-ring.
The constants $C_1$ and $C_2$ of Eq. (\ref{omega}) and (\ref{e_gamma}) 
are fixed as 
\[
C_1=\frac{\,\,2\sigma^{1/2}\,\al\,\,}{1+\al\beta},\ \ \ 
C_2=\frac{1}{\sqrt{2}(1+\al\beta)^2},
\]
to assure that the spacetime does not have global rotation
and that the periods of $\phi$ and $\psi$ 
become $2\pi$ at the infinity, respectively.

To make the metric component $g_{\phi\phi}$ be regular, we have to set the integration constants $\alpha$ and $\beta$ as
\begin{eqnarray}
\alpha =  \pm \sqrt{\frac{2(\delta_1-1)(1-\eta_2)}{(\lambda -1)(\delta_2-1)(1-\eta_1)}}, ~~
 \beta =  \pm \sqrt{\frac{(\lambda +1)(\delta_2+1)(1+\eta_1)}{2(\delta_1+1)(1+\eta_2)}} .
\end{eqnarray}
These conditions also assure the non-existence of closed timelike curves 
around the event horizons.

To cure the conical singularities, we have to set the periods of angular
coordinates appropriately.  
The periods of the coordinates $\phi$ and $\psi$ are defined as 
\begin{equation}
 \Delta \phi = 2 \pi \lim_{\rho \rightarrow 0} \sqrt{\frac{\rho^2 g_{\rho\rho}}{g_{\phi\phi}}} 
 ~~~\mbox{and}~~~
 \Delta \psi = 2 \pi \lim_{\rho \rightarrow 0} \sqrt{\frac{\rho^2 g_{\rho\rho}}{g_{\psi\psi}}}.
\end{equation}
We already set the periods of $\phi$ and the one 
of $\psi$ outside the ring to be $2\pi$. 
In addition the periods of $\psi$ can be obtained from
\begin{equation}
 \Delta \psi = \frac{2\pi}{1+\alpha\beta}
              \sqrt{\frac{(\lambda +1)(\lambda-\delta_2)}
                         {(\lambda - 1)(\lambda-\delta_1)}}
              \left(\frac{\lambda -\eta_2}{\lambda -\eta_1}\right)
              \left(1+\frac{\lambda - 1}{\lambda +1}\alpha\beta\right),
 \label{eq:psi_in}
\end{equation}
for $\delta_2\sigma<z<\lambda\sigma$ and
\begin{eqnarray}
 \Delta \psi &=& \frac{2\pi}{1+\alpha\beta}
              \sqrt{\frac{(\lambda +1)(\delta_1-1)(\delta_2+1)
                            (\delta_1-\eta_2)(\delta_2-\eta_1)}
                         {(\lambda - 1)(\delta_1+1)(\delta_2-1)
                            (\delta_1-\eta_1)(\delta_2-\eta_2)}} \nonumber \\&&
          \times    \left(\frac{\lambda -\eta_2}{\lambda -\eta_1}\right)
              \left(1+\frac{(\lambda - 1)(\delta_1+1))(\delta_2-1)}
                  {(\lambda +1)(\delta_1-1))(\delta_2+1)}\alpha\beta\right),
 \label{eq:psi_out}
\end{eqnarray}
for $\eta_2\sigma<z<\delta_1\sigma$. The parameters can be adjusted to make the both
values of $\Delta \psi$ equal $2\pi$.

Asymptotic form of $\mE_{S}$ near the infinity $\tilde{r}=\infty$ becomes 
\begin{eqnarray}
\mE_{S}&=&\tilde{r}\cos\theta\,
\left[\,1\,-\,\frac{\sigma}{\tilde{r}^2}\,\frac{P(\al,\beta,\lam)}
                                       {(1+\alpha\beta)^2}
     \,+\cdots\right] \nonumber  \\ &&  \hskip -0cm   
     +2\,i\,\sigma^{1/2}\,\left[\,\frac{\alpha}{1+\alpha\beta}
      \,-\,\frac{2\sigma\cos^2\theta}{\tilde{r}^2}\,\frac{Q(\al,\beta,\lam)}
                                                 {(1+\alpha\beta)^3}
    \,\,+\cdots\,\right],
\nonumber 
\end{eqnarray}
where we introduced the new coordinates $\tilde{r}$ and $\theta$ through
the relations
\footnote{In the previous version, one of these relations is inappropiate. As a result of this revision, the mass parameter is improved to be consistent with the ADM mass obtained by Yazadjiev, arXiv:0805.1600 [hep-th].}
\begin{equation}
 x=\frac{\tilde{r}^2}{2\sigma}+\lambda+(\eta_1-\eta_2)+(\delta_1-\delta_2),~~
 y=\cos 2\theta,
\end{equation}
and 
\begin{eqnarray}
P &=&  4(1 + \alpha^2 - \alpha^2 \beta^2) +2 (1+\alpha\beta)^2 (\delta_2 -\delta_1),     \nonumber  \\
Q &=& \alpha(2\alpha^2 -\delta_1+\delta_2+\eta_1-\eta_2+\lam+3)
-2\alpha^2\beta^3  \nonumber\\
    && 
  -\beta\left[2(2\alpha\beta+1)(\alpha^2+1) 
  +(\delta_1-\delta_2-\eta_1+\eta_2-\lam-1)\al^2(\al\beta+2)\right].  \nonumber
\end{eqnarray}
 From the asymptotic behavior of the Ernst potential, 
 we can compute the mass
parameter $m^2$ and rotational parameter $m^2a_0$ as
\begin{eqnarray}
&& m^2 = \sigma \frac{P}{(1+\alpha\beta)^2}, \hspace{0.2cm}
m^2a_0 = 4\sigma^{3/2}\frac{Q}{(1+\alpha\beta)^3}. \nonumber 
\end{eqnarray}

The angular velocities of event horizons are obtained from the direction vectors 
of finite timelike rods.
For the finite timelike rod of inner ring $[\delta_1\sigma<z<\delta_2\sigma]$, the
direction vector is calculated as
\begin{equation}
{\bf v}=(1,\Omega_1,0), ~~~
 \Omega_1=-\frac{2\beta(1+\alpha\beta)}
            {\sqrt{\sigma}((\lambda-1)\alpha\beta + \lambda+1)
                          ((\delta_2-1)\alpha\beta + \delta_2+1)}.
\end{equation}
The outer ring $[\eta_1\sigma<z<\eta_2\sigma]$ has a direction vector
\begin{eqnarray}
&& {\bf v}=(1,\Omega_2,0), \nonumber\\
&& \Omega_2=\frac{(1+\alpha\beta)((2\beta(\delta_1+1)(1+\eta_2)
                                -\alpha(\lambda+1)(\delta_2+1)(1-\eta_1))}
            {2\sqrt{\sigma}(2\alpha\beta(\delta_1+1)(1+\eta_2)
                           -(\lambda+1)(\delta_2+1)(\alpha^2(1-\eta_1)
             +2\alpha\beta+2))}.
\end{eqnarray}
When $\eta_1=-1$, the inner ring becomes static because of $\Omega_1=0$.
In this case the rotation of the outer ring only can cause the absence of the 
conical singularity. 
When $\eta_2=1$, the both rings rotate along the same direction.
 
Analyzing the mass and angular momentum parameters, we can show the infinite
nonuniqueness of black di-ring
which means that the di-ring solution has a continuous 
parameter region to have the same mass 
and angular momentum. 
In addition there can be two different single ring limits, thin and fat black rings,
of the black di-ring with the same mass and angular momentum.
Therefore these two single rings can be transformed into each other through
the black di-ring of the same mass and angular momentum.

\begin{figure}
  \includegraphics[scale=1.,angle=0]{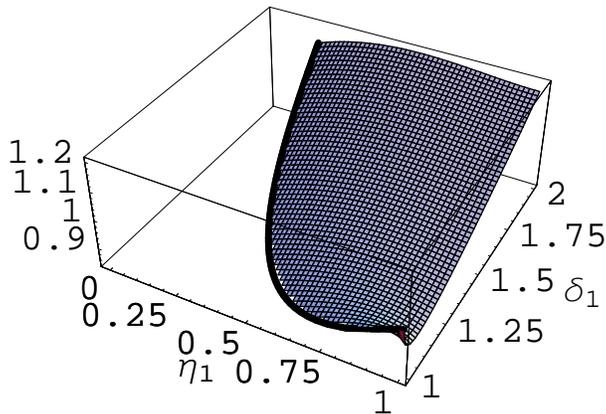}
  \caption{Plot of $\frac{a_0^2}{m^2}$ as a function of $\eta_1$ and $\delta_1$
           where $\lambda$ and $\delta_2$ are determined by the equilibrium 
           conditions and $\eta_2=1$. The bold line corespond to the single 
           black ring of $\delta_1=\delta_2$.}
 \label{fig:a02m2_3D}
 \end{figure}

\begin{figure}
  \includegraphics[scale=1.,angle=0]{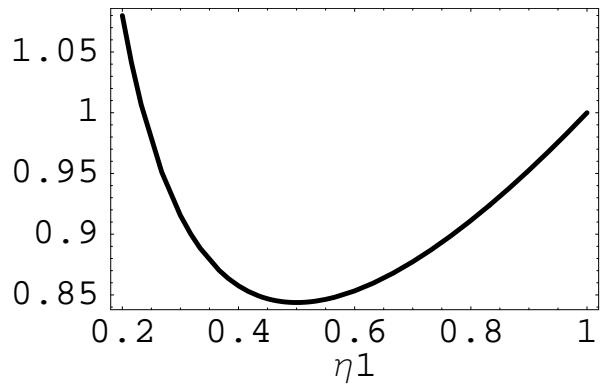}
  \caption{Plot of $\frac{a_0^2}{m^2}$ of single black ring 
           as a function of $\eta_1$. The region $0.5<\eta_1<1$ correspond to
           fat rings and $\eta_1<0.5$ thin rings.}
 \label{fig:a02m2}
 \end{figure}

To show this fact,
we consider the black di-ring of $\eta_2=1$. In Fig. \ref{fig:a02m2_3D}, we plot
the variable
\begin{equation}
 \frac{a_0^2}{m^2} = \frac{16 Q^2}{P^3}
\label{eq:a0m}
\end{equation}
as a function of $\delta_1$ and $\eta_1$.
At first we numerically decide the values of $\lambda$ and $\delta_2$ 
for the balanced black
di-ring for which the right hand sides of
Eqs. (\ref{eq:psi_in}) and (\ref{eq:psi_out})
become $2\pi$ with respect to given $\delta_1$ and $\eta_1$.
Next we obtain the value of $\frac{a_0^2}{m^2}$ 
by substituting the parameters which satisfy the condition
$\delta_1<\delta_2<\lambda$. 
The bold line of Fig. \ref{fig:a02m2_3D} is the single ring limit where
 $\delta_1=\delta_2$.
When $\delta_1=\delta_2$ we can show that Eq. (\ref{eq:a0m}) of equilibrium ring
is reduced to the following form,
\begin{equation}
 \frac{a_0^2}{m^2}=\frac{(1+\eta_1)^3}{8\eta_1},
 \label{eq:a0m_single}
\end{equation}
which corresponds with the single $S^1$-rotating black ring.
The Figure \ref{fig:a02m2} is a plot of Eq. (\ref{eq:a0m_single}), 
where $0.5<\eta_1<1$ corresponds to the fat ring and $\eta_1<0.5$ the thin ring.
Along the bold line in Fig. \ref{fig:a02m2_3D}, the value of 
$\frac{a_0^2}{m^2}$ has
the same $\eta_1$ dependence of Eq. (\ref{eq:a0m_single}).
Apparently there is a continuous path of di-ring between fat and thin 
black rings which have the same mass and angular momentum.

\section{Summay and discussion}
\label{sec:summary}

In this paper we have shown 
that  the $S^1$-rotating black ring can be superposed concentrically.
The solution of this
multiple rings can be written down 
by the solitonic transformation for the appropriately arranged seed solution.
We have obtained the functions needed to write down the metric of black di-ring
which is the simplest multiple $S^1$-rotating black ring.
To regularize the metric function, the integration constants $\alpha$ and 
$\beta$ should be set appropriately. 
For the equilibrium black di-ring we need the two additional
conditions of parameters. We have analyzed the mass and angular momentum of
black di-ring from the asymptotic form of Ernst potential.

The most important feature of black di-rings is that they entail 
the infinite nonuniqueness
of the vacuum neutral solutions of five dimensional general relativity.
To show this, we have numerically plotted 
the spin parameter of the equilibrium black di-rings
as a function of the two independent 
parameters. 
This plot shows that
there are infinite numbers of black di-rings 
with the same mass and angular momentum.
In addition we have shown that
the black di-ring can
be a pathway between the fat and thin $S^1$-rotating black rings.

The nonuniqueness we have shown is derived from the existence
of one-parameter family of black di-rings with
the same conserved parameters
because we have fixed one parameter in the analysis.
The parameters set for which the general black di-rings have
the same conserved parameters
would be a two-dimensional surface in the three-dimensional parameters space.
The physical features of black di-ring will be analyzed in detail.

The generalization of the solution to have two angular momenta
would be important. Recently, the generalization of 
the single black ring solution to this direction has been considered 
by the inverse scattering method \cite{Pomeransky:2006bd}
and by the numerical study \cite{Kudoh:2006xd}.
After this work was completed we noticed a preprint \cite{Elvang:2007rd},
which considers a black saturn: a spherical black hole surrounded by a black ring.
It would be important to consider the relation between the black saturn and the
black di-ring solutions.

\acknowledgments
We are grateful to T. Harmark for helpful discussions.
This work is partially supported by Grant-in-Aid for Young Scientists (B)
(No. 17740152) from Japanese Ministry of Education, Science,
Sports, and Culture.


\end{document}